\documentclass[12pt,a4paper]{article}
\usepackage{epsfig}
\begin{document}

\begin{center}
{\bf The critical points of lattice QCD with a non--zero quark density}
\\
Maria-Paola Lombardo 
\\{\it Zentrum f\"ur interdisziplin\"are Forschung,
  Universit\"at Bielefeld, \\  D-33615 Bielefeld, Germany}   
\end{center}
\begin{abstract}We study the interplay of
quark number density and chiral symmetry  in lattice QCD.
We suggest that both are controlled by the eigenvalue spectrum
of the fermionic propagator matrix, which shapes the
pattern of zeros of the partition functions.
The onset in the quark current would be triggered by the 
lowest lying eigenvalue, the chiral transition
 by the density of zeros, the two
critical points being distinct in full QCD, and coincident in the
quenched approximation. Our preliminary estimate for 
the critical point in full QCD in the infinite couling limit compares
favourably with the predictions of the strong coupling expansions
and of numerical simulations based on
exact, alternative representations of the partition function.
Several reasons of perplexity however remains, which are briefly
discussed.
\end{abstract}

Lattice QCD at non--zero quark number[1] is a poorly 
understood subject, despite the success of numerous
calculations exploiting approximation schemes or 
simpler models.
Many of the difficulties 
come from the particular structure of the Lagrangean:
as the building blocks of the 
lattice QCD Lagrangean
are quarks fields, a finite density on the lattice is realized by 
a coupling to a quark density, as opposed to a nucleon 
density of nuclear physics models. 
In practical numerical work, this induces several 
technical problems, which are described in detail in past publications. 
From a phenomenological point of view, 
it is not obvious that an excess of 3N quarks
would produce the same physical effects as an excess of N baryons
and it might be useful to keep this in mind,
especially when facing unexpected  results[2].

This warning issued, the rest of this note will  only deal with
the current
formulation of lattice QCD at non--zero quark number. The
numerical analysis  poses specific problems, since the action
is complex. The only method  which at the moment has the potential
to deal with it,  proposed by Barbour 
some years ago,  is based on  the representation of the partition
function $Z$ as 
$
Z = < Det (P - exp(\mu)> $
where $P$ is the fermionic propagator matrix[3].

The quark number density is an interesting observable which 
can be easily computed within this approach. 
The common feature of the results  for the number 
density [4,5], 
as reviewed by Ukawa
at this meeting, 
is a rather early onset $\mu_o$, definitively smaller than the critical
point for chiral symmetry restoration
expected around $m_N/3$ and a saturation
threshold, beyond which the particle density is one. 
$\mu_o$ can be rather accurately measured in the strong
coupling limit, where we find $\mu_o \simeq m_\pi/2$. 
High statistics
simulations  confirm that also at intermediate
coupling the number density is sensitive to the pion mass [5].

In the following we only discuss 
the subcritical region at zero temperature in the
infinite coupling limit, where the results they can be contrasted
with the predictions of the strong coupling expansions [6]
and of an alternative, exact representation of the
partition function [7].

We do not discuss here the  problems connected with the saturation 
threshold, which hampers the observation of the Stefan-Boltzmann
behaviour deep in the hot and dense phase, we just mention that  
possible solutions might be found in the framework
of the lattice improved/perfect actions discussed here by 
Karsch, Wiese and T.~D.~ Lee. 

\vskip .5 truecm
\noindent
{\it The early onset : the density of states of the fermionic
operator and the pion mass.}
Gibbs
proved that the onset of the number density 
on isolated configurations is controlled by the lowest
eigenvalue of the fermion propagator matrix, and argued
that the lowest eigenvalues is half the pion mass [8].
The results reviewed above suggest that this 
holds true also in the ensemble average. 
Alternative scenarios
can be proposed and we postpone their discussion to
a lengthier presentation.
Here we merely sketch an argument which suggests
that the persistency of this
result in the statistical ensemble at zero temperature
is compatible with the
symmetries of the system, so ergodicity
problems, if any, are not obviously manifest.

Consider the determinant on a isolated configuration, 
and the partition function after averaging over the statistical ensemble:
$Det = \prod_{i=1,6V} ( z-\lambda_i)$, 
$Z = \prod_{i=1,6V}( z - \alpha_i)$.
The {$\lambda_i$}'s are the eigenvalues of the 
fermionic propagator matrix,
$z$ is the fugacity $e^\mu$ and
the {$\alpha_i$}'s the zeros of the partition function in
the complex fugacity plane: the zeros of the partition function 
can be seen as the ``proper'' ensemble average
of the eigenmodes of the fermionic propagator matrix.
From the determinant we obtain the number density
on isolated configurations at zero temperature [8]:
$J_0 = 1/V \sum_{1 < |\lambda_i| < e^\mu} 1.$, and we 
note that an analogous expression holds true in the
ensemble average as well, provided that we trade the eigenvalues 
with the zeros of the partition function. 
To know the fate of the onset of the current in the statistical ensemble
we only need to monitor   min \{$|\lambda| 's$ \},  the contribution 
of each pole to the current  being,  configuration
by configuration, constant. Consider now that
the Z3 symmetry, well verified in high statistics
simulations [5], imposes $Z = \prod_{i=1,2V}( z^3 - \beta_i)$,
i.e. the Z3 symmetry constraints the arguments of the
zeros, but not their modulus. 
We have indeed checked that, configuration by configuration,
 min \{$\ln |\lambda|$ \} $\simeq m_\pi/2$ :it may well be that
the onset in the current in the full ensemble is the same as the
onset on isolated configurations, whose origin is clear.

In a sense, this is a  straighforward 
result: it says that the signal in the fermion number
density is triggered by the
lowest eigenmode in the  spectrum of the fermionic propagator matrix.
This result is the one expected 
for non confining theories, like Gross Neveu, where the lowest state defines
indeed the dynamical fermion mass [9].
In 
{\it this formulation and within this approach} 
to lattice QCD the lowest eigenvalue gives half
the pion mass. 

\vskip .5 truecm
\noindent
{\it The chiral transition, and random matrix models}
The possibility of observing another phase transition at 
$\mu_c > \mu_o $, and a non trivial distribution of
zeros of the partition function are closely related, as we
can read off the expression for the number density.
How would these transitions at $\mu_o$ 
and $\mu_c$ relate with chiral symmetry restoration?

A very useful laboratory for the study of chiral symmetry
is offered by random matrix models. In the case of QCD at finite
density, we learnt from the work of Stephanov
that the  transition would show at a
physical $\mu_c$ in the full model, but at half the pion mass in 
the quenched approximation [10]. 
The applications of these results to QCD would
suggest that the onset for the quark number at half the pion
mass would also restore the chiral symmetry in the quenched model,
 because of the simultaneous occurrence
of quarks and conjugate quarks in the system, while 
in  full QCD the  chiral symmetry, would be restored at the ``correct''
 $\mu_c$ [6,7] thanks to the
rearrangements of the eigenvalues produced by the richer dynamics--
this relates also to
the different  nature of critical phenomena in quenched and full models.
\footnote{
Given the prominent role of the density of states
evaluated at zero $\mu$, the suggestion would naturally arise
to give a second try to various partial quenched schemes [11].
These approximations were dismissed in the past because
of the threshold at half the pion mass, but, as we have
shown, this threshold is not necessarily related with the chiral transition.
Particularly interesting could be their application 
 to the recent
proposal by Kogut and collaborators [12].}

This discussion suggests a natural numerical strategy,  whose
preliminary outcome is shown in Figure 1.  These
results , obtained on a $6^4$ and $8^4$ lattice with
a bare quark mass = .1, are preliminary, and are just meant 
as an illustration of the simple idea presented above.
 We see that  the two statistical
ensembles (quenched and full) show the same extrema, which defines a
common critical region for the quenched and the full model.
This agrees with the results of [13], where the ``forbidden''
region of quenched QCD was found to be coincident with the metastable
region of full QCD, as computed in strong coupling expansion.
However, contrary to the expectations of [13], it seems impossible 
to measure the real chiral transition point 
in the quenched approximation, while in the full
model a peak in the eigenvalues distribution shows up in correspondence
with the expected critical point.
\footnote{The is reminiscent of other observations:
 quenched and full model can be  completely different, 
but even dramatic qualitative differences are realized by subtle numerical
effects.}
Needless to say, it would be very
interesting to study the relation of the spectrum of the fermionic
propagator matrix with the fermion matrix spectrum, which is naturally
related with chiral symmetry.

\begin{figure}
{\epsfig{file=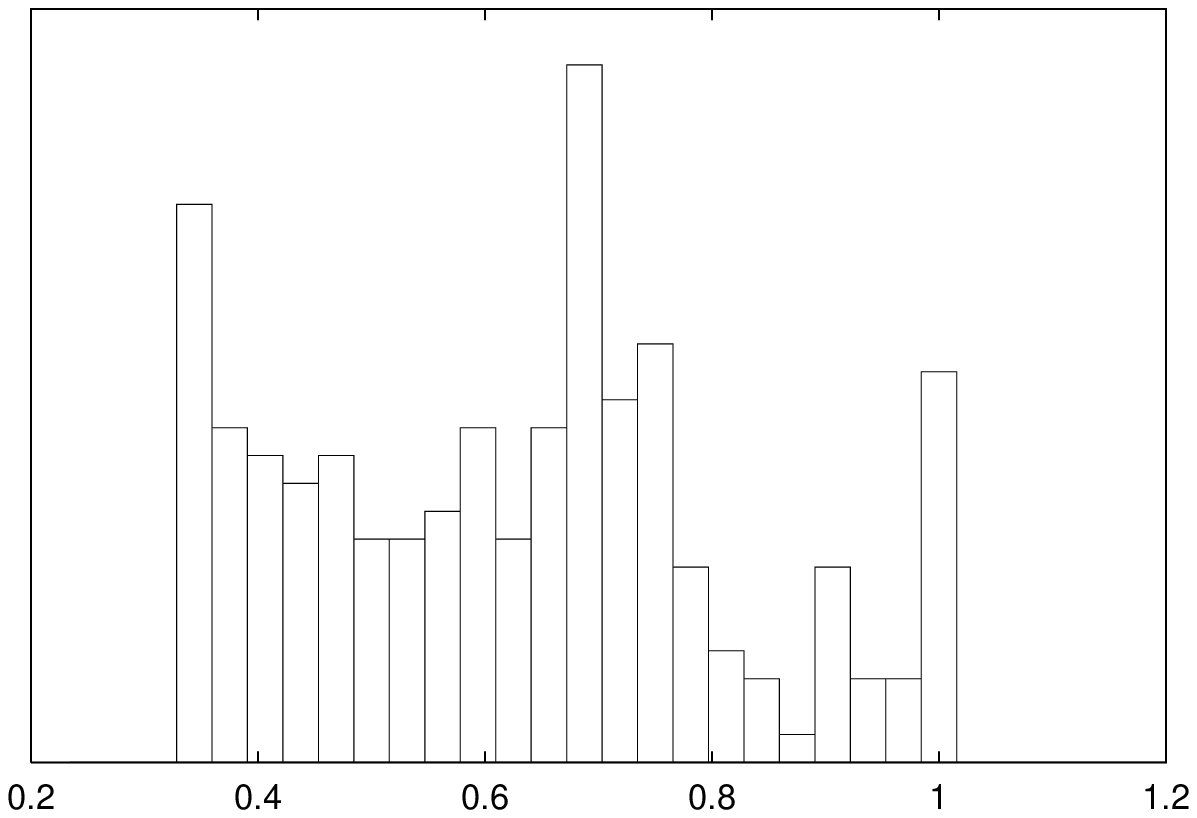, width=7cm, angle=0} 
\epsfig{file= 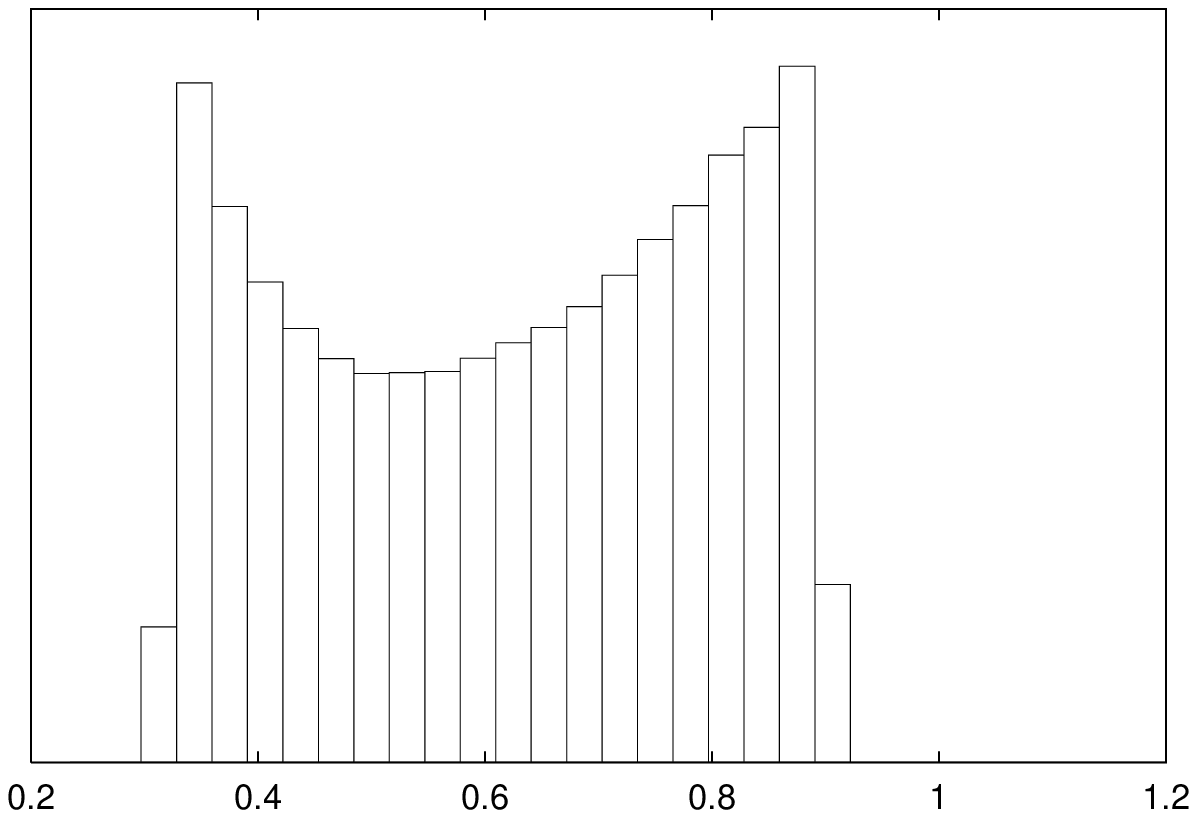, width=7cm, angle=0}}
  \hspace*{3cm} $\ln |\alpha|$ \hspace*{6cm} $\ln | \lambda|$ 
  \caption {Density of states for the grand 
canonical partition function
(left) and the fermionic propagator matrix, which is its quenched 
counterpart:  the threshold, compatible with $m_\pi/2 = .31$[13]  induces 
a non zero number density in the quenched and full model alike;
one appealing scenario could be that
this threshold induces the restoration of chiral symmetry in the
quenched model, but not in the full model where
the chiral transition point,  estimated from the
the position of the peak , is consistent with
$\mu_c = .69(1)$[7]}
\end{figure}


\vskip .5 truecm
\noindent
{\it Summary}
The picture suggested by the above
discussions is as follows : on isolated configurations the analysis of
the fermionic propagator matrix gives a signal
for the current in correspondence to 
its lowest state.
This seems to survive the statistical 
ensmble average: there are two main reasons for this, first, that
the real part of the pole is stable, second, 
that for each pole the amplitude of the contribution 
to the current is constant, and equal 1, so no cancellation occurs.
Thus the onset for the quark number is the lowest real part
of the zero of the partition function, or, equivalently,
the real part of the logarithm of the
lowest eigenmode of the fermionic propagator matrix.
This a straighforward result in non--confining models,  
while in QCD the result $\mu_o \simeq m_\pi/2$, and the
nature of the region $\mu_o < \mu < \mu_c$ ,  
deserve further investigation.

We haven't found any indication of systematic errors 
associated with the method or lack of ergodicity in the
algorithm which can offer an alternative explanations of
these observations. Still, they cannot be exluded. The best way
to address this issue, in our opinion, is to cross check with
the results of alternative formulations$^{12}$.

Within this formulation  quenched and 
full QCD  share the same critical region. However, 
the simple dynamics of the quenched  model cannot build up
any structure in the density of modes, so no new transition
appears after statistical averaging, and the results
from random matrix models suggest that the onset of the current 
restores chiral symmetry.
The structure in the  spectrum is
instead apparent in  full QCD, where we attempted, not
unsuccesfully,  an estimate of the chiral transition point.

These observations would predict a non--zero critical density at the
zero temperature chiral transition,
and might be related with the presence of diquarks in the region
$\mu_o < \mu < \mu_c$. But, again,  cross checks with
other formulations are mandatory in order to disentangle possible
numerical artifacts from predictions amenable to an experimental
verification.

\vskip .3 truecm
This note reports on work in progress 
with I.~Barbour, S.~E.~Morrison, E.~G.~Klepfish 
and J.~B.~Kogut. I wish to thank my collaborators
and acknowledge valuable discussions with
F. Karsch. I would like to thank the
Physics Department of the University of Bielefeld for its
hospitality,  and the High Energy Group of HLRZ/J\"ulich,
particularly K. Schilling, for support during the initial
stages of this project. 
The calculations were done at  HLRZ/KFA J\"ulich.
This work was partially supported by Nato  grant no. CRG 950896.

\vskip .5 truecm
\noindent
\small {\it References}

\noindent
1. J.~B.~Kogut, H.~Matsuoka, M.~Stone, H.~W.~Wyld,
  S.~Shenker, J.~Shigemitsu, D.~K.~Sinclair, 
Nucl. Phys. {\bf B225} [FS9], (1983) 93;
 P.~Hasenfratz and F.~Karsch, Phys. Lett. 
{\bf125B} (1983) 308. \\
2. An interesting criticism of the current formulation can be inferred
from the  paper by D.~ Kharzeev, Phys.~Lett. {\bf B378} (1996) 238.\\
3. I.~Barbour, Nucl.~Phys.~ {\bf B} (Proc. Suppl.) 29 
(1992) 22.\\
4. A.~Hasenfratz and D.~Toussaint, Nucl. Phys. {\bf B371}
(1992) 539.\\
5. I.~Barbour, J.~Kogut and S.~E.~Morrison {\it Phase
structure of lattice QCD at finite density with dynamical fermion},
talk given by S.~E.~Morrison at Lattice96, 14th International
Symposium on Lattice Field Theory, St. Louis, MO, 4-8 June 1996,
hep-lat/ 9608057.\\
6. N.~Bili\'c, 
K.~Demeterfi and B.~Petersson, Nucl. Phys.{\bf B377} (1992) 615.\\
7. F.~Karsch and K.H. M\"utter, Nucl. Phys. {\bf B313}
  (1989) 541.\\
8. P.E. Gibbs, Phys. Lett. {\bf 172B} (1986) 53. \\
9. S.~ Hands, S.~ Kim and J.B.~Kogut, 
 Nucl.Phys. {\bf B442} (1995) 364.\\
10. M.~I. Stephanov, Phys. Rev. Lett. 76 (1996) 4472.\\
11.  J.~Engels and H.~Satz, Phys. Lett. {\bf 159B}
(1985) 151. \\  
12. J.~B.~Kogut and D.~K.~Sinclair, {\it QCD with chiral
4--fermion interaction}, talk given by D.~K. Sinclair at Lattice96,
hep-lat/9607083;
 I. Barbour, J.~B.~Kogut and S.~E.~Morrison, {\it QCD at finite
density}, talk given by I. Barbour at the International
Conference {\it Multi Scale Phenomena
and their Simulation}, 
Zentrum f\"ur interdisziplin\"are Forschung,Bielefeld, Germany,
30 September-- 4 October 1996.\\
13. M.-P. Lombardo, J.~B.~Kogut and D.~K.~ Sinclair,
Phys. Rev. {\bf D54} (1996) 2303.
\end{document}